\documentclass[aps, prl, 10pt, twocolumn, longbibliography]{revtex4-1}

\usepackage{textcase}
\usepackage{braket}
\usepackage{amsmath}
\usepackage{amssymb}
\usepackage[utf8]{inputenc}
\setcitestyle{square}
\usepackage{color}
\usepackage{graphicx}% Include figure files
\usepackage{dcolumn}% Align table columns on decimal point
\usepackage{bm}% bold math
\usepackage[colorlinks=true, citecolor=blue, linkcolor=blue]{hyperref}

\begin{document}

\preprint{APS/123-QED}

\title{Dynamical order and superconductivity in a frustrated many-body system}

\author{J. Tindall$^{1}$, F. Schlawin$^{1}$, M. Buzzi$^{2}$, D. Nicoletti$^{2}$, J. R. Coulthard$^{1}$, H. Gao$^{1}$, A. Cavalleri$^{1,2}$, M. A. Sentef$^{2,3}$ and D. Jaksch$^{1,4}$}
\affiliation{$^1$Clarendon Laboratory, University of Oxford, Parks Road, Oxford OX1 3PU, United Kingdom} 
\affiliation{$^2$ Max Planck Institute for the Structure and Dynamics of Matter, 22761 Hamburg, Germany}
\affiliation{$^3$Institute for Theoretical Physics, University of Bremen, Otto-Hahn-Allee 1, 28359 Bremen, Germany}
\affiliation{$^4$Centre for Quantum Technologies, National University of Singapore, 3 Science Drive 2, Singapore 117543}

\date{\today}

\begin{abstract}
In triangular lattice structures, spatial anisotropy and frustration can lead to rich equilibrium phase diagrams with regions containing complex, highly entangled states of matter. In this work we study the driven two-rung triangular Hubbard model and evolve these states out of equilibrium, observing how the interplay between the driving and the initial state unexpectedly shuts down the particle-hole excitation pathway. This restriction, which symmetry arguments fail to predict, dictates the transient dynamics of the system, causing the available particle-hole degrees of freedom to manifest uniform long-range order. We discuss implications of our results for a recent experiment on photo-induced superconductivity in ${\rm \kappa - (BEDT-TTF)_{2}Cu[N(CN)_{2}]Br}$ molecules. 
\end{abstract}

\maketitle

%\tableofcontents

\par \textit{Introduction --} Identifying and understanding the processes which prevent thermalization and decoherence in driven-dissipative quantum systems \cite{Sieberer_Dynamical_2013} is a unifying theme in ultracold atoms and condensed matter research \cite{Brennecke_Real-time_2013,Abanin_Effective_2017}. This comes with the potential to realize and functionalize exotic out-of-equilibrium quantum phases, both for the continued progress of fundamental research and for wider technological purposes. In ultrafast materials science, the counterintuitive experimental observation of light-induced superconductivity \cite{OpticalSCExperiment1, OpticalSCExperiment2, OpticalSCExperiment3, OpticalSCExperiment4, OpticalSCExperiment5, OpticalSCExperiment6, OpticalSCExperiment7, KappaSaltExperiment, OpticalSCExperiment8} has stimulated the field. In these experiments intense laser pulses have been reported to induce superconducting-like features, such as an inverse-frequency divergence of the imaginary part of the optical conductivity and vanishing resistivity, well above the materials' equilibrium critical temperatures, $T_{c}$. 
\par In a very recent experiment, specific vibrational modes of the charge-transfer salt ${\rm \kappa - (BEDT-TTF)_{2}Cu[N(CN)_{2}]Br}$ were resonantly excited with mid-infrared radiation and the above-mentioned optical features were induced at temperatures several times higher than $T_{c}$ \cite{KappaSaltExperiment}. Moreover, following excitation, a large gap in the real part of the optical conductivity opened up -- a feature not seen when cooling the molecular crystal below $T_{c}$. These results suggest a different mechanism for superconductivity compared to that when cooling the material. Within Ref.~\cite{KappaSaltExperiment}, a minimal microscopic two-rung triangular Hubbard lattice, with time-dependent parameters under resonant driving of specific phonon modes, was proposed as a model for the experiment.
\par A number of theoretical studies have explored the effects of carefully-tuned coherent driving on the prethermal dynamics of one and two dimensional bi-partite fermionic lattice models \cite{SCTheory1, SCTheory2, SCTheory3, SCTheory4, SCTheory5, Coulthard, Sentef1, Murakami2017}. These studies are motivated by the opportunities arising from having dynamical time-dependent Hubbard parameters, which have been experimentally realised in contexts ranging from quantum simulators \cite{Messer_Floquet_2018} to strongly-correlated materials via electronic  \cite{Tancogne-Dejean_Ultrafast_2018, Ishikawa2014, Wall2011} as well as vibrational excitations \cite{Singla_THz-2015}. Their relevance, however, to organic materials such as the ${\rm \kappa-(BEDT-TTF)_{2}}X$ compounds is unclear, due to the dimerized ${\rm BEDT-TTF}$ molecules forming a half-filled triangular, non bi-partite lattice \cite{KappaModels1, KappaModels2, KappaModels3, KappaModels4}. Instances of the triangular Hubbard model, alongside other non-bipartite Hubbard lattices, do not possess the same symmetries as their hypercubic counterparts and the frustration and hopping anisotropy can lead to rich equilibrium phase diagrams containing unique states of matter \cite{TriangularPhaseTransition1, TriangularPhaseTransition2}.

\par In this paper we demonstrate how the interplay between such equilibrium states and generic periodic driving manifests complex nonequilibrium behavior in a triangular Hubbard model. Motivated by the results of Ref.~\onlinecite{KappaSaltExperiment}, and the opportunity to explore the many-body dynamics of a driven frustrated system, we consider the time-dependent two-rung triangular Hubbard model and identify two distinct phases when driving the ground state out of equilibrium. Beneath a critical value of the vertical hopping integral $\tau' < \tau'_{c}$ there is a unique phase where the particle-hole excitation pathway is unexpectedly blocked. This impedance, which symmetry arguments fail to predict, causes the driving to establish amplified, coherent, long-range particle-hole order in the available degrees of freedom. As $\tau'$ increases across the critical value, $\tau'_{c}$, this restriction in the particle-hole channel is lifted and the driven system cannot dynamically sustain order due to the creation of a number of incommensurate particle-hole excitations.   

\begin{figure}[t]
\centering
\includegraphics[width = \columnwidth]{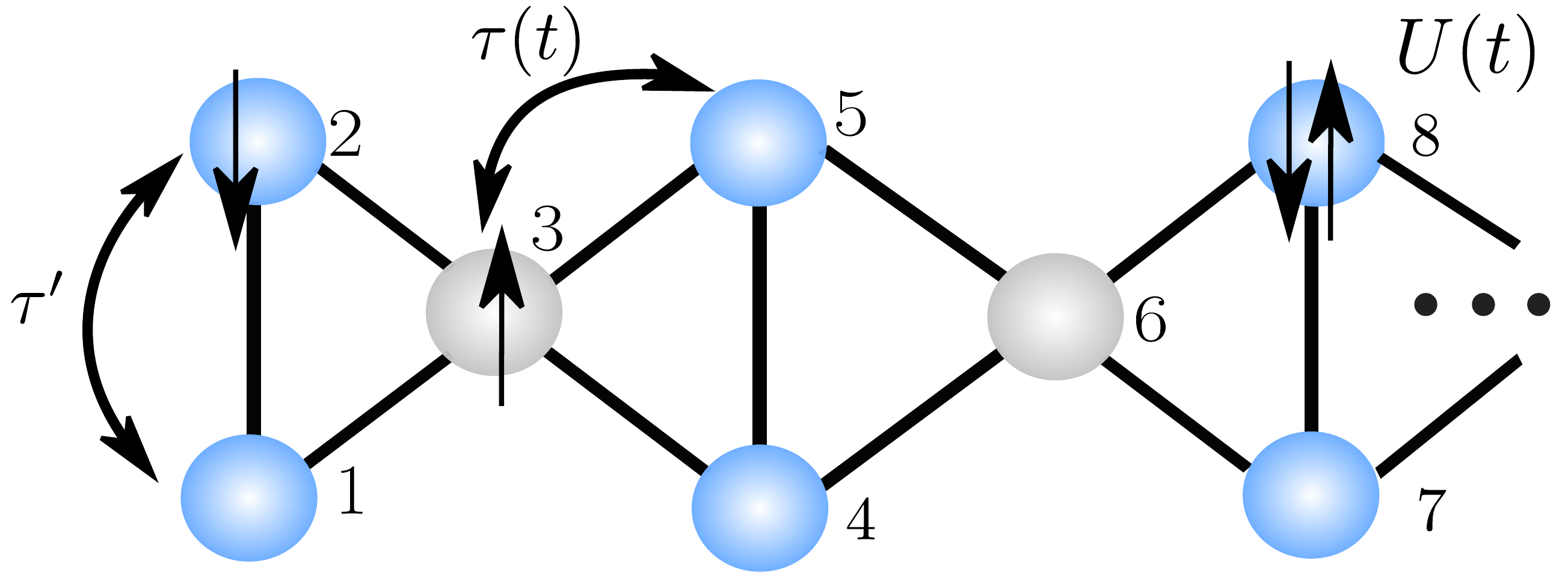}
\caption{First $8$ sites of the two-rung triangular Hubbard model described by Eq.~(\ref{Eq:KappaHam}). The model has a time-dependent nearest-neighbor hopping $\tau(t)$, static vertical hopping $\tau'$, and a time-dependent local Hubbard interaction $U(t)$. The outer (blue) vs central (grey) sites represent a bi-partite splitting of the lattice in the limit $\tau' = 0$.}
\label{Fig:F1}
\end{figure}

\par We proceed to identify the origin of these distinct regimes - a rich ground state phase diagram with properties not seen in hypercubic realisations of the Hubbard model. When $\tau' < \tau'_{c}$, the system forms a spin-wave condensate with an extensive interference pattern induced by the anisotropic geometry of the system. The condensed nature of this initial state prevents the driving from causing excitations in the particle-hole SU(2) pathway and leads to the observed induction of long-range order. For $\tau' > \tau'_{c}$ this spin-wave condensate is depleted and the driving produces excitations which instigate the decay of any particle-hole correlations.
\par Finally we show how, even for a small amplitude pulse, driving the spin-condensed initial state near resonance causes rapid relaxation towards a doublon-ordered state. The parameters we use reflect the dynamical electronic properties of the photo-excited ${\rm \kappa - (BEDT-TTF)_{2}Cu[N(CN)_{2}]Br}$ molecules in Ref. \cite{KappaSaltExperiment}. We thus offer a possible explanation for the physical mechanism that underlies the transient onset of superconductivity observed in this experiment. More broadly, our results provide an understanding of how geometrical effects can significantly alter the non-equilibrium behavior of driven systems.
\par \textit{Model and Method --} The Hubbard model is a paradigmatic quantum lattice model which has relevance for high-temperature superconductivity \cite{HubbardSC}, can be realized in ultracold atom experiments \cite{HubbardExperiment}, and is solvable using the Bethe ansatz in one dimension \cite{HubbardBethe, HubbardSymmetries, Shastry}. The rich symmetry structure of the model is responsible for this solubility. On a bi-partite lattice, there are two SU(2) symmetries, known as the `spin' and `$\eta$' symmetries \cite{Yang_Eta_1989}, which play a significant role in the physics of the model. For example, driving/dissipative terms which preserve the $\eta$ symmetry have been shown to guide the system into steady states with long-range correlations in the $\eta$ channel, a phenomenon named heating-induced order \cite{HeatingInducedOrder}.  
\par Here we focus on the role of heating in the dynamics of a driven non-bipartite two-rung triangular Hubbard model, where the $\eta$ symmetry does not exist. The Hamiltonian is
\begin{align}
H(t) = -\tau(t) \sum_{ij \in \langle {\rm n. n} \rangle, \sigma}(c^{\dagger}_{\sigma, i}c_{\sigma, j} + {\rm h.c}) \ - \notag \\ \tau' \sum_{ij \in \langle {\rm vert } \rangle, \sigma}(c^{\dagger}_{\sigma, i}c_{\sigma,j} + {\rm h.c}) + U(t)\sum_{i}n_{i, \uparrow}n_{i, \downarrow},
\label{Eq:KappaHam}
\end{align}
where $n_{\sigma, i}$, $c_{\sigma, i}^{\dagger}$ and $c_{\sigma, i}$ are, respectively, number, creation and annihilation operators for fermions of spin $\sigma \in \{\uparrow, \downarrow\}$ on site $i$. In Eq.~(\ref{Eq:KappaHam}), the first summation is a time-dependent hopping term, with strength $\tau(t)$, over the diagonal nearest-neighbor bonds pictured in Fig.~\ref{Fig:F1}. The second summation is the hopping term $\tau' H_{V}$ over the vertical bonds and the last term is a time-dependent interaction term, with strength $U(t)$. We consider a half-filled lattice with $L$ sites and a total magnetisation of $0$.
\par The time-dependence of the nearest-neighbour hopping and interaction strengths is
\begin{align}
&\tau(t)  = \bar{\tau}\big(1 + A_{\tau}\sin^{2}(\Omega t)\exp \big(-(t - T_{p})^{2}/(2T_{w}^{2})\big)\big),  \notag \\
& U(t)  = \bar{U} \big(1+A_{U}\sin^{2}(\Omega t)\exp \big(-(t - T_{p})^{2}/(2T_{w}^{2})\big)\big),
\label{Eq:Pulses}
\end{align}
which constitutes a fairly general parametrisation of the Hamiltonian parameters. In Eq. (\ref{Eq:Pulses}), $A_{U}$ and $A_{\tau}$ are the amplitudes of the modulation of $U$ and $\tau$ relative to their equilibrium values $\bar{U}$ and $\bar{\tau}$. The frequency of the oscillations is $\Omega$, whilst $T_{p}$ and $T_{w}$ describe the offset and width of the Gaussian envelope containing these oscillations. Our observations in this paper are not specific to the parameters of the driving. In the Supplemental Material (SM) we demonstrate our results choosing different parameters to those in the main text \cite{SM}.

\begin{figure*}
\centering
\includegraphics[width = \textwidth]{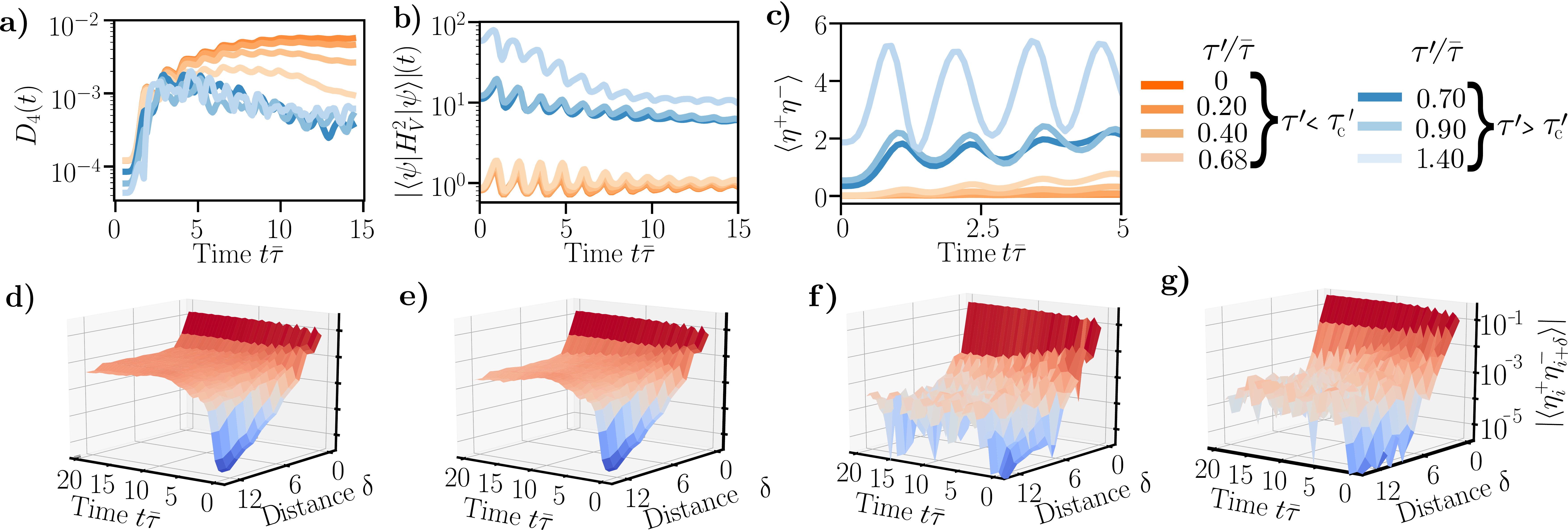}
\caption{Dynamics of the half-filled $L=14$-site two-rung triangular Hubbard model. The system is initialized in the ground state of $H(0)$, setting $\bar{U} = 5.0\bar{\tau}$ with the specified $\tau'$ and time-evolved under $H(t)$, using the same $\bar{U}$ and $\tau'$, with $A_{U} = -0.75, \ A_{\tau} = A_{U}/2$, $\Omega = 2.5\bar{\tau}$, $T_{p} = 0$ and $T_{w} = \infty$. a) Long-range doublon order $D_{4}(t)$ versus time, where $D_{4}(t)$ is defined in Eq (\ref{Eq:DoublonOrder}). b) Evolution of $\langle H_{V}^{2} \rangle$, where $\tau'H_{V}$ corresponds to the second hopping term in Eq. (\ref{Eq:KappaHam}). c) Evolution of $\langle \eta^{+}\eta^{-} \rangle$. d-g) Time and distance dynamics of the particle-hole correlations for $\tau' = 0.2\bar{\tau}, 0.4\bar{\tau}, 0.9\bar{\tau}, 1.4\bar{\tau}$ respectively.}
\label{Fig:F2}
\end{figure*} 

%\par By modelling the dynamics of the Hamiltonian in Eq. (\ref{Eq:KappaHam}) we are able to explore the non-equilibrium behaviour of a triangular, geometrically frustrated lattice structure. Furthermore, by using a two-rung ladder we can adapt Matrix Product State (MPS) methods and perform time-evolution calculations outside the remit of exact diagonalisation.

\par We start by identifying the symmetry structure of $H(t)$. Firstly, we show that there is a permanent spin SU(2) symmetry  $[H(t), S^{\pm,z}] \equiv 0 $ where $S^{\pm}$ and $S^{z}$ are the total spin raising/lowering and counting operators respectively \cite{HubbardSymmetries}.
We also find
\begin{equation}
[H(t), \eta^{z}] \equiv 0, \quad [H(t), \eta^{+}\eta^{-}] = [\tau' H_{V}, \eta^{+}\eta^{-}] \propto \tau',
\label{Eq:EtaSymmetry}
\end{equation}
where
\begin{equation}
\eta^{+} = \sum_{i = 1}^{L}f(i)c_{i, \uparrow}^{\dagger}c_{i, \downarrow}^{\dagger}, \qquad \eta^{z} =  \sum_{i}(n_{\uparrow, i} + n_{\downarrow, i} - 1),
\label{Eq:EtaOperators}
\end{equation}
and $\eta^{-} = (\eta^{+})^{\dagger}$ are the total $\eta$ operators and act on the doublons (locally paired fermions) and holons (empty sites) within the lattice. In Equation (\ref{Eq:EtaOperators}), $f(i)$ takes the value +1(-1) for the blue (grey) lattice sites in Fig.~\ref{Fig:F1}.
Equation (\ref{Eq:EtaSymmetry}) reveals that for finite $\tau'$ the system does not possess an $\eta$ ${\rm SU}(2)$ symmetry due to the presence of the vertical hopping term $H_{V}$. We emphasize that even in the bi-partite limit, $\tau' \rightarrow 0$, the system is not equivalent to the 1D Hubbard model due to the differing co-ordination numbers on the blue vs grey sites.

\par With this knowledge of the symmetries in hand, we extend the methodology of \cite{LongTimeDriving, HighFreq1, PONTE2015196} and propose that, in the long-time limit of $H(t)$, the system will reach a state of maximum entropy subject to the constraint that expectation values of conserved quantities must be preserved. If $H(t)$ possesses an SU(2) symmetry, this constraint leads to heating-induced order, with the long-time state guaranteed to have uniform, long-range correlations in the conserved symmetry sector \cite{Buca2019, HeatingInducedOrder}. Meanwhile, for the symmetries that $H(t)$ does not preserve, a large number of incommensurate excitations will be created - causing the decay of correlations in the corresponding sectors \cite{HeatingInducedOrder}.  Hence we expect that for finite $\tau'$ large amplitude driving from Eq. (\ref{Eq:Pulses}) should cause the particle-hole correlations to quickly decay away due to the lack of an $\eta$ SU(2) symmetry for $H(t)$.
\par \textit{Results --} In the following we investigate this, initializing the system in the ground state of $H(0)$ and time-evolving it under $H(t)$. We quantify the correlations in the $\eta$ symmetry sector using the particle-hole function $\vert \langle \eta^{+}_{i}\eta^{-}_{j} \rangle (t) \vert$ which describes the mobility of a doublon between sites $i$ and $j$ at time $t$.
We also introduce the doublon order parameter
\begin{equation}
    D_{\delta}(t) = (1/N)\sum_{\substack{ij \\ |i-j| \geq \delta}} \vert \langle \eta^{+}_{i}\eta^{-}_{j} \rangle (t) \vert,
    \label{Eq:DoublonOrder}
\end{equation}
where $N$ is a constant such that $D_{\delta}(t)$ is the average of the particle-hole function for distances greater than $\delta - 1$.
\par In Fig.~\ref{Fig:F2} we drive the system with a long, large-amplitude pulse for different values of $\tau'$.  We observe that for $\tau' \neq 0$, $\langle \eta^{+}\eta^{-} \rangle$ is not conserved as expected.  The plots in Fig. \ref{Fig:F2}, however, reveal that there is a critical value of $\tau'$ where the behavior of the system in the particle-hole channel changes significantly under driving. For $\tau' < \tau'_{c} \approx 0.69\bar{\tau}$ uniform, doublon order forms on transient time-scales and follows closely the $\tau' = 0$ evolution, despite the absence of the requisite symmetry. Meanwhile, for $\tau' > \tau'_{c}$ the system's response in the particle-hole sector is much less ordered and the corresponding off-diagonal correlations quickly decay away. This distinct change in the system's behaviour is underpinned by the action of the vertical hopping term $H_{V}$. In Figs. \ref{Fig:F2}b-c we see that for $\tau' < \tau'_{c}$ this term effectively acts like an annihilation operator which shuts down the particle-hole excitation pathway, preventing $\langle \eta^{+}\eta^{-} \rangle$ from changing significantly and inducing long-range order amongst the available particle-hole correlations. As $\tau'$ increases above $\tau'_{c}$ this is no longer the case and the rate of change of $\langle \eta^{+}\eta^{-} \rangle$ jumps by over an order of magnitude due to the creation of incommensurate particle-hole excitations which prevent robust order from being established in this sector.

\par
\begin{figure}[t]
\centering
\includegraphics[width = \columnwidth]{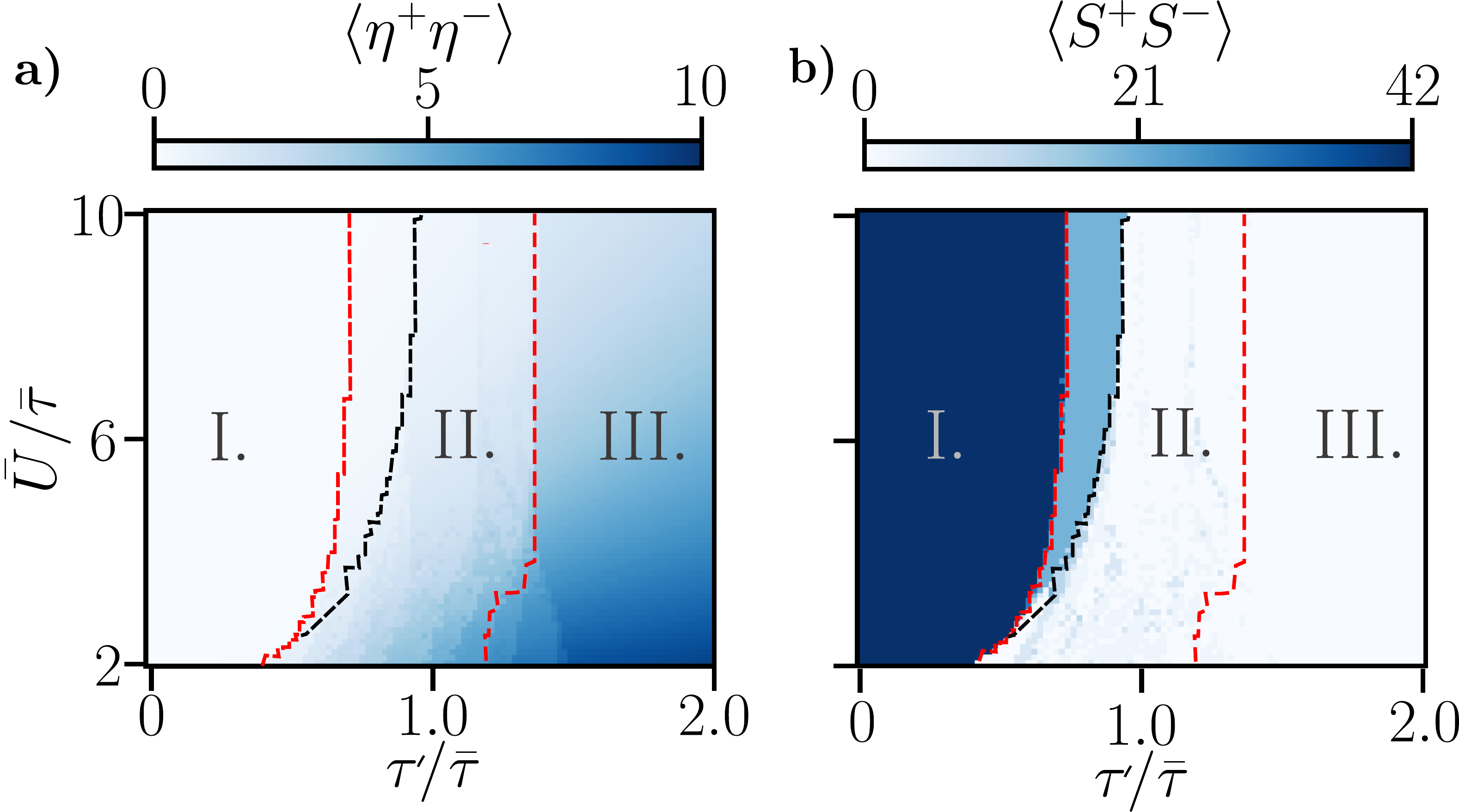}
\caption{a-b) Map of $\langle \eta^{+}\eta^{-} \rangle$ and $ \langle S^{+}S^{-} \rangle$ versus $\bar{U}$ and $\tau'$ for the ground state of the $L=32$-site two-rung triangular Hubbard model. The \textit{red} dotted lines separate the three distinct phases/regions - ${\rm I}$, ${\rm II}$ and ${\rm III}$ - observed for this system size, their properties are described in the main text. In the thermodynamic limit the width of region ${\rm I}$ changes and the ${\rm I}$ - ${\rm II}$ transition instead occurs along the \textit{black} dotted line.}
\label{Fig:F3}
\end{figure}

 We now probe the origin of these two distinct phases, calculating the properties of the ground state of the system which we drove out of equilibrium. In Fig.~\ref{Fig:F3} we observe a rich phase diagram for the ground state of $H(0)$ in terms of $\langle \eta^{+}\eta^{-}\rangle$ and $\langle S^{+}S^{-} \rangle$. We also provide plots of the two-point correlations for states within these diagrams, alongside a finite-size scaling analysis (FSA) which indicates the phases we observe persist in the thermodynamic limit \cite{SM}.

Within region ${\rm I}$ of Fig.~\ref{Fig:F3}a the ground state of the system resides in the lowest eigenspace of $\langle \eta^{+}\eta^{-} \rangle$. Here, the system displays the properties of a spin-wave condensate through the large value of  $\langle S^{+}S^{-} \rangle$ underpinned by long-range spin-exchange order and a sharp $0$ momentum peak in the corresponding structure factor. There are, however, two additional peaks of opposite momenta which correspond to interference in the condensate order due to further correlations between the central and outer sites of the lattice. This Spin-Wave Condensate (SWC) arises from the the irregular geometry of the lattice and is not possible in hypercubic realisations of the half-filled Hubbard model, where the ground state always has $\langle S^{+}S^{-} \rangle = 0$. 
\par As $\tau'$ increases, the value of the condensate order parameter $\langle S^{+}S^{-} \rangle$ jumps discontinuously and the system undergoes a first-order phase transition into region ${\rm II}$. In this phase, some of the vertically-bonded sites localise and form singlets separate from the rest of the system, in which only small signatures of the condensate order observed in the previous phase remain. For even higher $\tau'$ the system undergoes another first order phase transition into region ${\rm III}$. Here, the vertical hopping is sufficiently large to create a spin-dimerized phase where all the vertically-bonded sites form singlets decoupled from the central sites. 
\par For finite-size systems the ${\rm I}$ - ${\rm II}$ transition occurs along the left-most red dotted line in Fig. \ref{Fig:F3}, where the condensate order parameter $\langle S^{+}S^{-}\rangle$ first jumps discontinuously. This is consistent with the critical change in non-equilibrium behaviour observed for $\tau' \approx 0.69\bar{\tau}$ in Fig. \ref{Fig:F2}. Our FSA indicates, however, that in the thermodynamic limit this transition, which is of first-order, occurs along the black-dotted line instead. This is because, as system size increases, the properties of the system in the light-blue region in Fig. \ref{Fig:F3}b converge to those of phase ${\rm I}$. We thus anticipate that the critical change in the non-equilibrium dynamics that we observed in Fig. \ref{Fig:F2} will occur at the higher value of $\tau'_{c} \approx 0.86\bar{\tau} $ when $L \rightarrow \infty$.
\par We understand this critical change in the non-equilibrium dynamics through the action of the vertical hopping term $H_{V}$ in the different phases. In region ${\rm I}$ $H_{V}$ acts as an annihilation operator, leading to the approximate conservation of $\langle \eta^{+}\eta^{-}\rangle$ and the formation of doublon-holon order. This action can be understood from the spin-wave nature of the ground state and that $H_{V}$ only acts in the sub-lattice containing the outer sites of the system (marked in blue in Fig. \ref{Fig:F1}). In this sub-lattice the spin-exchange correlations are large, positive and completely uniform with distance and so the two-site reduced density matrix (RDM) will be dominated by terms such as $(\ket{\uparrow \downarrow} + \ket{\downarrow \uparrow})(\bra{\uparrow \downarrow} + \bra{\downarrow \uparrow})$; which are annihilated by a Hubbard hopping operator. Moreover, on short timescales the driving acts mainly to modify the longer-range correlations in the system, transiently preserving the form of the two-site RDM on neighbouring sites and thus the action of $H_{V}$ as an annihilation operator. Meanwhile, in phases ${\rm II}$ and ${\rm III}$, the condensate order disappears and the vertically bonded sites begin to form singlets. A hopping term will map a singlet onto an orthogonal state and thus $H_{V}$ has a significant effect when acting on the ground state. In the SM we reinforce these statements by plotting the two-site RDM in all $3$ regions and computing the square-norm of $H_{V}$ for the full phase diagram in Fig. \ref{Fig:F3}. We also plot the resulting dynamical conservation, or lack thereof, of $\langle \eta^{+}\eta^{-} \rangle$ for various different driving parameters \cite{SM}.

\begin{figure}[t]
\centering
\includegraphics[width = \columnwidth]{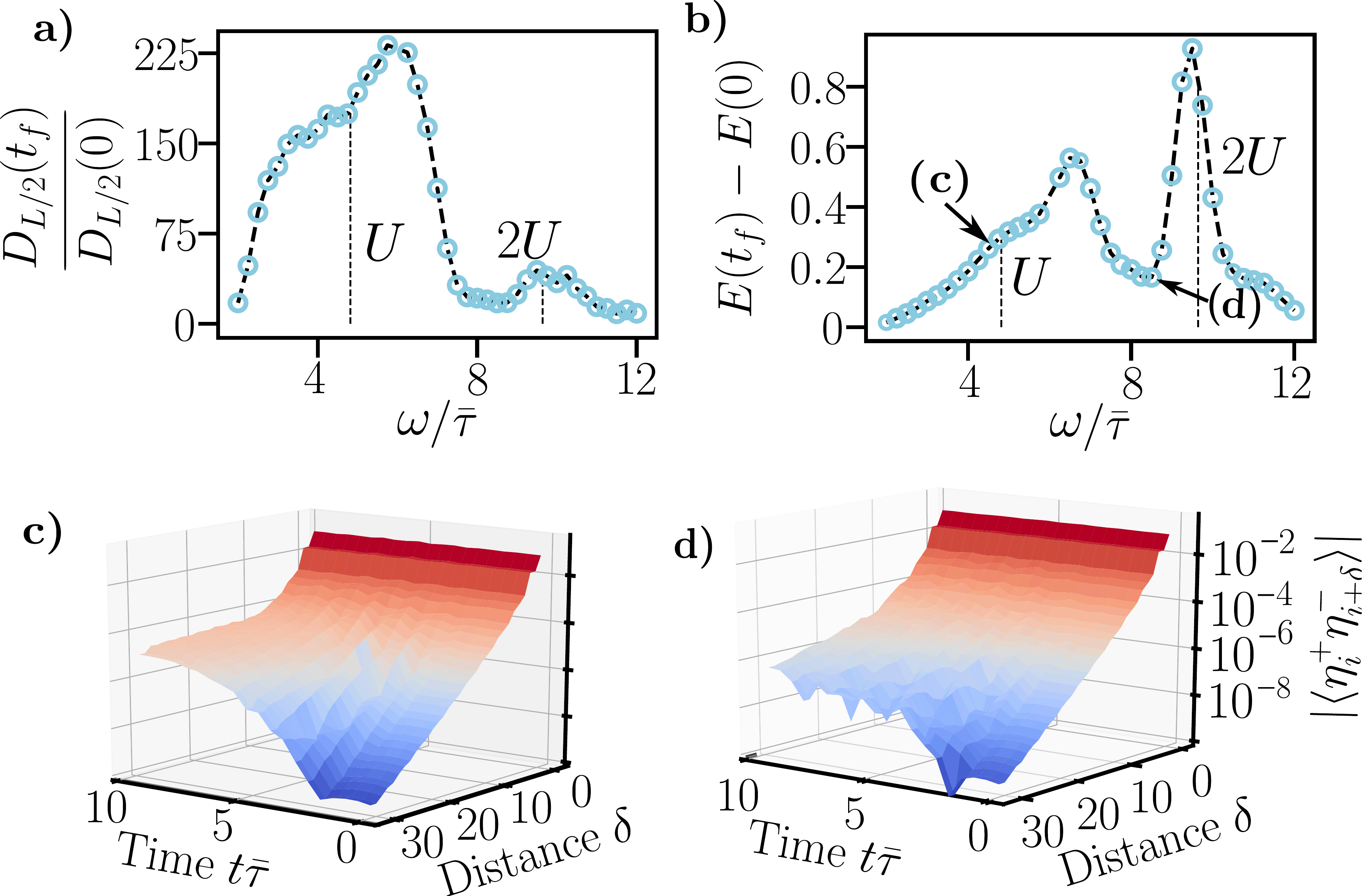}
\caption{Dynamical properties of the half-filled $L=32$-site two-rung triangular Hubbard model. The system is initialized in the ground state of $H(0)$, with $\bar{U} = 4.8 \bar{\tau}$ and $\tau' = 0.25 \tau$, and evolved under $H(t)$ with the same $\bar{U}$ and $\tau'$ whilst $A_{U} = -0.15$, $A_{\tau} = A_{U}/2$, $T_{p} = 5.0\bar{\tau}$, $T_{w} = 2.5\bar{\tau}$ with the specified $\omega = 2\Omega$. a) Ratio of the long-range doublon order $D_{L/2}(t)$, see Eq. (\ref{Eq:DoublonOrder}), at times $t_{f}\bar{\tau} = 10$ and $t = 0$ versus $\omega$. b) Energy  difference between the initial state and that at time $t_{f}\bar{\tau} = 10.0$ (where $E(t) = \langle H(0) \rangle (t)$). c-d) Time and distance dynamics of $\vert \langle \eta^{+}_{i}\eta^{-}_{i + \delta} \rangle \vert$ for $\omega = 4.5 \bar{\tau}$ and $\omega = 8.5 \bar{\tau}$, these frequencies are marked in b). The parameters used to create (c) reflect the electronic changes induced by the driving of ${\rm \kappa - (BEDT-TTF)_{2}Cu[N(CN)_{2}]Br}$ in Ref. \cite{KappaSaltExperiment}.}
\label{Fig:F4}
\end{figure}

\par In Figure \ref{Fig:F2} we drove the system with a long, large amplitude pulse, demonstrating the formation of stable doublon order for $\tau' < \tau'_{c}$. In Fig.~\ref{Fig:F4} we consider a larger system and use a shorter-lived, smaller amplitude pulse to drive the system out of equilibrium for $\tau' < \tau'_{c}$, showing that the annihilative action of $H_{V}$ still dictates the dynamics of the system. We focus on the role of the driving frequency, plotting the long-range order and absorbed energy as a function of the `bare' frequency $\omega = 2\Omega$ (the pulse form is $\sin^{2}(\Omega t))$ in Figs.~\ref{Fig:F4}a-b. These quantities both display doubly peaked profiles, with the peaks in the doublon order coinciding with those of the absorbed energy. This is indicative of heating-induced order, with the system relaxing to an ordered state due to the approximate conservation of $\langle \eta^{+}\eta^{-} \rangle$ as it gains energy \cite{HeatingInducedOrder}. 
\par We further illustrate this by showing the explicit evolution of the particle-hole correlations for two different frequencies in Figs. \ref{Fig:F4}c-d. There is a slower induction of order at the higher frequency, with the system also absorbing less energy from the driving field. Despite the finite value of $\tau'$ we see the system is transiently relaxing towards a state with amplified, uniform off-diagonal correlations.
\par Interestingly, in Figs. \ref{Fig:F4}a-b, for $\omega \approx 2U$ the system absorbs the most energy and yet less long-range doublon order is induced compared to the first peak. This first peak is broad and shifted from $\omega = U$, which is consistent with the pulse causing a number of excitations which all evolve differently and are necessary for the successful re-arrangement of the particle-hole degrees of freedom. The second peak is of a sharp, non-dispersive nature, containing only a small number of excitations which remain localised during the time evolution. Here, the time dynamics has become diabatic: the driving frequency is the dominant timescale and is too rapid for the system to significantly adapt its spatial configuration \cite{Coulthard, HighFreq1, SM}. 
\par \textit{Conclusion --} We have studied the dynamics of the anisotropic driven two-rung triangular Hubbard model, showing how the rich ground state properties interact with the driving to create a phase which symmetry arguments fail to predict. In this phase, particle-hole excitations are prevented, causing the system to relax towards a state with amplified, uniform, particle-hole correlations.
\par The choice of a triangular Hubbard model was partly motivated by the role it has played in the modelling of the ${\rm \kappa-(BEDT-TTF)_{2}}X$ compounds. The rich dynamical behaviour which occurs under driving may therefore be observable in these materials. Along these lines, the Hamiltonian in Eq. (\ref{Eq:KappaHam}) was proposed as a model for a recent experiment optically exciting the vibrational modes of ${\rm \kappa - (BEDT-TTF)_{2}Cu[N(CN)_{2}]Br}$ \cite{KappaSaltExperiment}. The manifestation of particle-hole order we have witnessed here provides a potential explanation for the observed onset of light-induced superconductivity in this experiment. The experimental parameters are consistent with the system being driven close to $\omega = U$ from within the SWC phase and, more specifically, the Hamiltonian parameters and driving terms used in Fig. 4c were based on frozen-phonon simulations which determined the electronic properties of the photo-excited ${\rm \kappa - (BEDT-TTF)_{2}Cu[N(CN)_{2}]Br}$ molecules in Ref. \cite{KappaSaltExperiment}.
\par More generally, the two-rung triangular Hubbard model allowed us to study the dynamics of a non-equilibrium, geometrically frustrated lattice structure. We anticipate that other systems which possess geometries characterised by frustration, inhomogeneous co-ordination numbers and anisotropic hopping terms - such as Kagom\'e lattices \cite{HexagonalHubbard}, optical quasi-crystalline structures \cite{QuasiCrystals}, and doped cuprates \cite{Cuprates} - will display similarly rich non-equilibrium physics. 
\newline 

\begin{acknowledgments}
\textit{Acknowledgements} - We would like to acknowledge B. Buca for helpful discussions. We performed our calculations using Matrix Product State methods adapted using the Tensor Network Theory library \cite{TNT}.
This work has been supported by EPSRC grants No. EP/P009565/1 and EP/K038311/1 and is partially funded by the European Research Council under the European Union’s
Seventh Framework Programme (FP7/2007-2013)/ERC Grant Agreement No. 319286 Q-MAC. MAS acknowledges support by the DFG through the Emmy Noether programme (SE 2558/2-1). JT is also supported by funding from Simon Harrison and DJ partially carried out this work while visiting the Institute for Mathematical Sciences, National University of Singapore in 2019. Finally, our calculations were run on
the University of Oxford Advanced Research Computing (ARC) facility http://dx.doi.org/10.5281/zenodo.22558. 
\end{acknowledgments}

\providecommand{\noopsort}[1]{}\providecommand{\singleletter}[1]{#1}%

\clearpage
\setcounter{equation}{0}
\renewcommand{\theequation}{S\arabic{equation}}

\onecolumngrid

\section{Supplementary Material To Dynamical Order and Superconductivity in a Frustrated Many-Body System - Figures at the end of the text}

\subsection{Equilibrium Phase diagram of the two-rung triangular Hubbard Hamiltonian}
Here we identify the key properties of the different phases of the Hamiltonian
\begin{align}
&H(t) = -\tau(t) \sum_{ij \in \langle n. n \rangle, \sigma}(c^{\dagger}_{\sigma, i}c_{\sigma, j} + {\rm h.c}) - \tau' \sum_{ij \in \langle {\rm vert } \rangle, \sigma}(c^{\dagger}_{\sigma, i}c_{\sigma,j} + {\rm h.c}) + U(t)\sum_{i}n_{i, \uparrow}n_{i, \downarrow}.
\label{Eq:SKappaHam}
\end{align}
A description of the various terms and the lattice geometry are provided in the main text. For convenience, we refer to the outer sites of the ladder as sub-lattice A and the central sites as sub-lattice B - these correspond to the blue and grey sites in Fig. \ref{Fig:SF1} (top left) respectively. We also define the vertical hopping term $H_{V}$ as 
\begin{equation}
    H_{V} = -\sum_{ij \in \langle {\rm vert } \rangle, \sigma}(c^{\dagger}_{\sigma, i}c_{\sigma,j} + {\rm h.c}).
    \label{Eq: SHV}
\end{equation}
\par Firstly, we explore the properties of the ground state of $H(0)$, with $\bar{U} = U(0)$ and $\bar{\tau} = \tau(0)$ being the equilibrium values of the nearest-neighbour hopping and interaction strengths. In the top row of Fig \ref{Fig:SF1}, for a range of $\bar{U}$ and $\tau'$, we reproduce the phase diagram from the main text as a function of $\langle S^{+}S^{-}\rangle$. We have also provided a diagram of the geometry of the system here, with the numbers indicating the manner in which sites are indexed. In the remainder of the figure we plot the spin-exchange and magnetic order, as well as the corresponding structure factors, for the $3$ different phases. The relevant operators for computing these matrices are
\begin{align}
S^{z}_{i} = n_{\uparrow, i} - n_{\downarrow, i}, \qquad S^{+}_{i}  =c_{i, \uparrow}^{\dagger} c_{i, \downarrow}, \qquad S^{-}_{i}  =c_{i, \downarrow}^{\dagger} c_{i, \uparrow},
\end{align}
and the corresponding  structure factors are defined as
\begin{align}
S_{\pm}(q) = \sum_{jk}e^{i(j-k)q}\langle S^{+}_{j}S^{-}_{k} \rangle, \qquad S_{z}(q) = \sum_{jk}e^{i(j-k)q}\langle S^{z}_{j}S^{z}_{k} \rangle,
\label{Eq:Struct}
\end{align}
where $q$ ranges in discrete steps of ($2\pi/L$) from $-\pi$ to $\pi$.

Within phase ${\rm I}$, the system is similar to that of a Spin-Wave Condensate (SWC). Here, there is long-range spin-exchange order throughout the system which does not decay with distance and this leads to the large value of $\langle S^{+}S^{-} \rangle$ within this phase. Within sub-lattice A (blue sites in top left of Fig. \ref{Fig:SF1}) this order is positive and completely uniform with distance. This order leads to a single structure factor peak at $q = 0$, which indicates the presence of a robust spin-wave condensate. When considering the whole system, however, there is a staggered pattern in the spin-exchange correlations and two additional peaks of opposite momenta in the structure factor appear. This shows there is an interference pattern in the spin-exchange order due to the correlations between the two sub-lattices. This interference co-exists with the condensate.
\par As $\tau'$ increases the system transitions into phase ${\rm II}$. Here, $\tau'$ is strong enough to break up the condensate and some of the vertically bonded pairs of sites become asymmetric singlets unbound from the rest of the lattice. This is indicated by the formation of a small dip at $\delta = 1$ in the magnetic and spin-exchange correlations in sub-lattice A. The remaining sites still retain some of the correlations seen in phase ${\rm I}$, however these decay with distance and the $0$ momenta peak in the spin-exchange structure factor has been depleted.
\par Finally, for large enough $\tau'$ the system resides in a spin-dimerized phase (phase ${\rm III}$) where all the vertically bonded sites form asymmetric singlets unentangled with the rest of the system. These can be seen in the sharp dip at $\delta = 1$ in both the spin-exchange and magnetic correlations. There is no long-range order in any part of the system, and all the fermions have become localised.

\par In Fig. \ref{Fig:SF2} we consider how the properties of these states changes with system size in order to classify the transitions observed in the phase diagram of Fig. \ref{Fig:SF1}. We restrict ourselves to $U = 5.0\bar{\tau}$ and note that the discontinuous behaviour we observe appears for a range of $U$, with the critical values of $\tau'$ dependent on the explicit value of $U$. For $U = 5.0\bar{\tau}$ we observe 3 key critical points at $\tau' \approx 0.69\bar{\tau}, 0.86\bar{\tau}$ and $1.35\bar{\tau}$, where discontinuities appear in a number of observables: the derivative of the energy per site with respect to $\tau'$, the average number of doublons in the system and the total normalised spin-exchange order $\langle S^{+}S^{-} \rangle /L^{2}$. These discontinuities appear for all the system sizes we consider and thus point to the possible existence of several first-order phase transitions.
\par As the system size increases we observe that the size of the first discontinuity at $\tau' \approx 0.69\bar{\tau}$ is diminishing towards $0$ whilst the other two appear to be stable/ increasing. Hence, in the thermodynamic limit the two points $0.86\bar{\tau}$ and $1.35\bar{\tau}$ correspond to first-order phase transitions between the different phases previously identified: for $\tau' < 0.86\bar{\tau}$ the system is in phase ${\rm I}$ consistent with the finite value of $\lim_{L \rightarrow \infty} \langle S^{+}S^{-} \rangle/(L^{2}) = 1/36$ which was obtained by finding exact polynomials for the scaling of $\langle S^{+}S^{-} \rangle$ with system size.
For $0.86\bar{\tau} <\tau' < 1.35\bar{\tau}$ the system is in phase ${\rm II}$ and for $\tau' > 1.35\bar{\tau}$ the system is in phase ${\rm III}$.
\par We note that the value of the transition between phases ${\rm I}$ and ${\rm II}$ in the thermodynamic limit is distinct from that observed in the finite size calculations. For example, for $L=14$ we observe this transition at $0.69\bar{\tau}$, where the condensate order first exhibits a discontinuity and we observe a critical change in the system's behaviour under driving (Fig. 2 of the main text). This distinction appears because as the system size increases the condensate order parameter in the region $0.69\bar{\tau} <\tau' < 0.86\bar{\tau}$ increases from $0$ (for $L= 8$) and approaches that of the phase ${\rm I}$. This region is approximately covered by the light blue area of the phase diagram in Fig. \ref{Fig:SF1}. This suggests that in the thermodynamic limit the critical change in the system's dynamics under driving should occur at $\tau' \approx 0.86\bar{\tau}$, where the condensate order jumps to $0$ (top right, Fig. \ref{Fig:SF2}). Hence, the driving can only induce long-range doublon-holon order for $\tau' < 0.86\bar{\tau}$. 

\subsection{Non-Equilibrium Behaviour of the two-rung triangular Hubbard Hamiltonian}

\par In Figs. \ref{Fig:SF3} and \ref{Fig:SF4} we concern ourselves with the effect of $H(t)$ when dynamically evolving the phases discussed in the previous section. Specifically, we focus on the role of the vertical hopping integral $H_{V}$ and how it affects the conservation of $\langle \eta^{+}\eta^{-} \rangle$ as the system absorbs energy from the driving. 
\par Firstly, in Fig. \ref{Fig:SF3}, we see that within the phase ${\rm I}$ $H_{V}$ effectively annihilates the ground state. This can be understood from its spin-wave condensed nature and the fact the vertical hopping integral acts between sites solely within sub-lattice A (blue sites in top right of Fig. \ref{Fig:SF1}). Within this sub-lattice the spin-exchange correlations are large, positive and completely uniform with distance and in Fig. \ref{Fig:SF3} we see that the two-site reduced density matrix (RDM) in this region is approximately of the form
\begin{align}
    \rho \approx \rho' = c_{1}(\ket{\uparrow, \uparrow}\bra{\uparrow, \uparrow} + \ket{\downarrow, \downarrow}\bra{\downarrow, \downarrow}) + c_{2}( \big(\ket{\uparrow, \downarrow} + \ket{\downarrow, \uparrow}\big)\big(\bra{\uparrow, \downarrow} + \bra{\downarrow, \uparrow}\big))
    \label{Eq:RDM}
\end{align}
where $c_{1}$ and $c_{2}$ are constants which are an order of magnitude larger than any of the other matrix elements in $\rho$. The term $c_{2} \big(\ket{\uparrow, \downarrow} + \ket{\downarrow, \uparrow}\big)\big(\bra{\uparrow, \downarrow} + \bra{\downarrow, \uparrow}\big)$ is the cause of the large spin-exchange correlations and it is straightforward to see that $\rho'$ (and thus approximately $\rho$) is annihilated by a Hubbard hopping operator acting from either the left or the right. Moreover, $\rho'$ is exactly that one obtains when taking the two-site RDM of the spin-wave condensate $(S^{+})^{X/2}\ket{\downarrow_{1} \downarrow_{2} ... \downarrow_{X}}$ for any lattice with $X$ sites. Hence, the annihilative action of $H_{V}$ in region ${\rm I}$ is a direct result of the spin-wave nature of the ground-state in sub-lattice A. 
\par Meanwhile in the other two phases the ground state is no longer annihilated by $H_{V}$. In fact we can see that from the reduced density matrix the vertically bonded sites form a state with strong asymmetric singlet correlations, which a hopping term will have a significant effect on (hopping terms map anti-symmetric singlets onto an orthogonal state with an identical norm).

\par These observations are reflected in Fig. \ref{Fig:SF4} where we time evolve the ground state under $H(t)$ for various different driving parameters. We see that $\langle \eta^{+}\eta^{-} \rangle$ is effectively unchanged in phase ${\rm I}$ which can be understood from the annihilative action of $H_{V}$ - this is no longer true in the other phases where $\langle \eta^{+}\eta^{-} \rangle$ is clearly not conserved. 
\par Although $H_{V}$ acts as an annihilation operator on the ground state and $\langle \eta^{+}\eta^{-} \rangle$ should therefore be conserved on a very short time-scale, one could argue that this will no longer be the case after a reasonable period of time as the driving will have caused a significant change in the initial wavefunction. This is not what we observe and in Fig. 2 of the main text we see that $H_{V}$ acts as an annihilation operator over fairly long periods of time, resulting in the approximate conservation of $\langle \eta^{+}\eta^{-} \rangle$ and the build up of doublon-holon order. We understand this from the fact that, transiently, driving at the frequencies and amplitudes we use in the main text mainly causes changes in the longer-range correlations in the system - as opposed to more local, nearest-neighbour correlations. This can be seen in Fig. \ref{Fig:SF5}, where we compare the effect of two different driving frequencies on the correlations in the system. As a result the two-site RDM for a vertically bonded pair of sites stays close to that observed in Fig. \ref{Fig:SF3} and $H_{V}$ transiently annihilates the non-equilibrium state when driving from within the phase ${\rm I}$.

\subsection{Diabatic Behaviour When Driving at Large Frequencies}
In Fig. \ref{Fig:SF5} we plot the dynamics of the particle-hole correlations at various distances for two different driving frequencies, with the form of the driving now the same as in the main text
\begin{align}
&\tau(t)  = \bar{\tau}\big(1 + A_{1}\sin^{2}(\Omega t)\exp \big(-(t - T_{p})^{2}/(2T_{w}^{2})\big)\big),  \notag \\
& U(t)  = \bar{U} \big(1+A_{2}\sin^{2}(\Omega t)\exp \big(-(t - T_{p})^{2}/(2T_{w}^{2})\big)\big).
\label{Eq:SPulses}
\end{align}
The two frequencies we consider correspond to the two peaks of the distributions in Fig. 4 of the main text. We see that at the higher frequency, the driving induces a much stronger change in the short-range correlations as opposed to the longer-range correlations. The opposite occurs at the lower frequency. This is a consequence of, at the higher frequency, the driving frequency being the largest timescale in the system. As a result a `freezing effect' occurs where the system struggles to adapt its spatial configuration and the driving induces a more local response in the system's dynamics \cite{Coulthard, HighFreq1}.

\setcounter{figure}{0}  
\renewcommand{\thefigure}{S\arabic{figure}}

\begin{figure}[h]
\centering
\includegraphics[width = 0.8\textwidth]{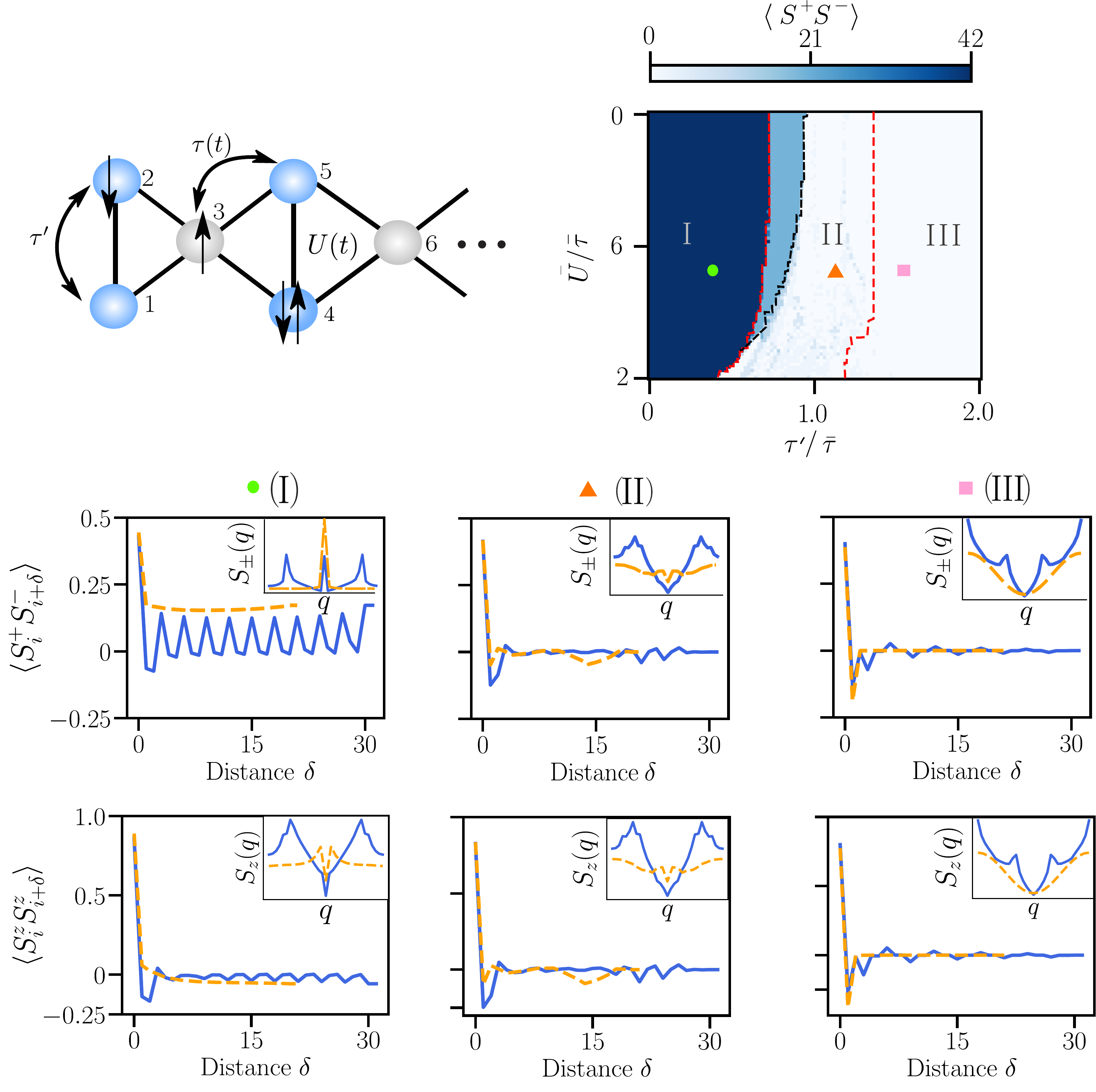}
\caption{Top Row) Reproduced from Fig.1 and Fig. 3b of the main text. Left - Geometry of the two-rung Hubbard model described by Eq. (\ref{Eq:SKappaHam}). The numbers indicate the indexing of the sites, with the blue and grey sites corresponding to sub-lattices A and B respectively. Right - Expectation value of $\langle S^{+}S^{-} \rangle$ for the ground state of the $L=32$ two-rung triangular Hubbard model, see Eq. (\ref{Eq:SKappaHam}), as a function of $\bar{U}$ and $\tau'$. The \textit{red} dotted lines separate the three distinct phases/regions - ${\rm I}$, ${\rm II}$ and ${\rm III}$ - observed for this system size, their properties are described in the text. In the thermodynamic limit the width of region ${\rm I}$ changes and the ${\rm I}$ - ${\rm II}$ transition instead occurs along the \textit{black} dotted line. Second and Third Rows) Respectively, spin-exchange and magnetic correlations versus distance $\delta$. Insets) Spin-exchange and magnetic structure factors, see Eq. (\ref{Eq:Struct}), versus quasi-momenta $q$. Orange-dotted lines show the corresponding quantities when calculated solely over sub-lattice A, which corresponds to the blue sites in the lattice depicted in the top left plot.}
\label{Fig:SF1}
\end{figure}

\begin{figure*}
\centering
\includegraphics[width = 0.8\textwidth]{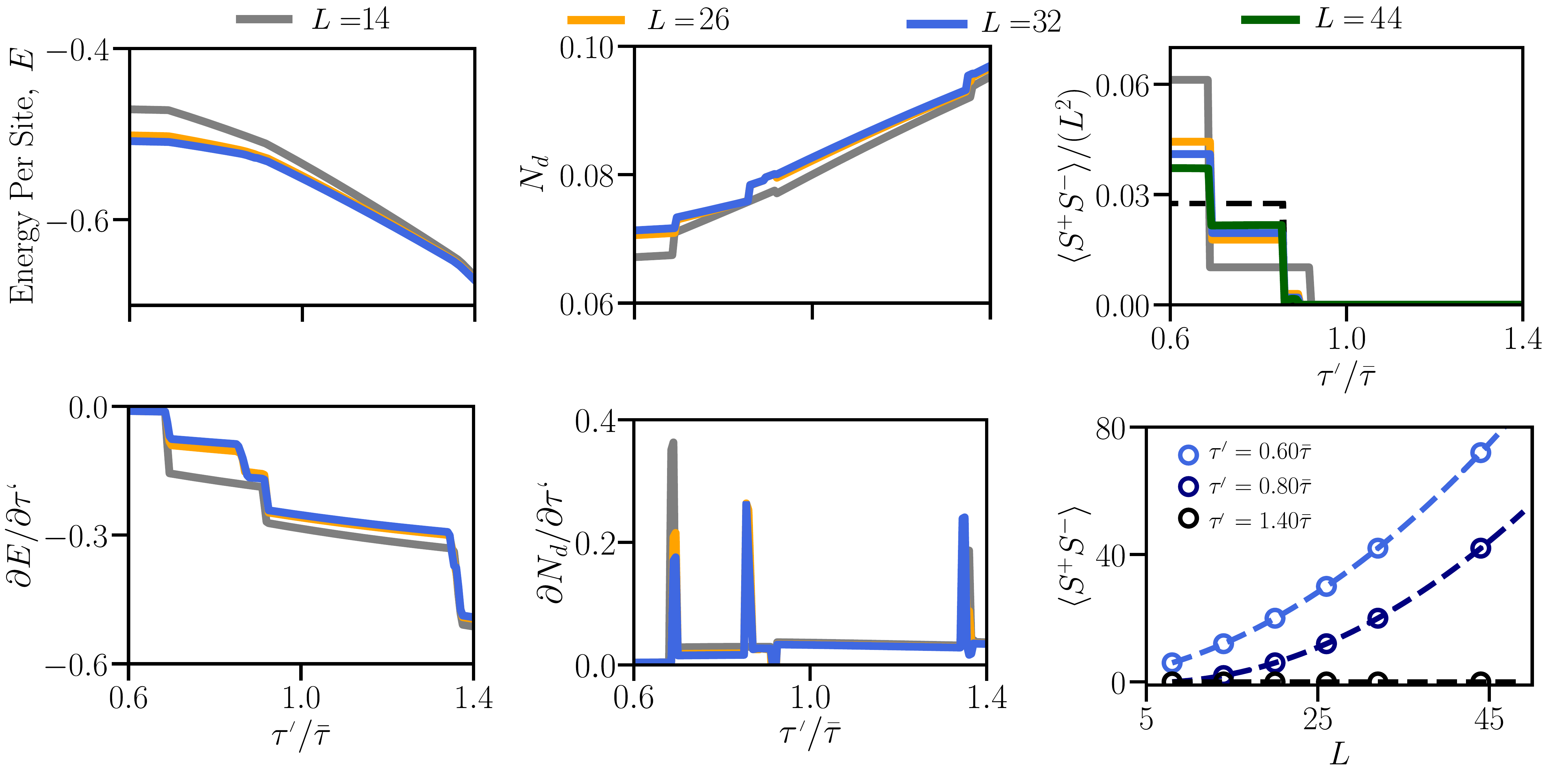}
\caption{Top Row) Energy per site, average doublon occupancy $N_{d}$ and spin-exchange order parameter verus $\tau'$ for the ground state of the two-rung triangular Hubbard model described in (Eq. \ref{Eq:SKappaHam})  with $U = 5.0\bar{\tau}$. Several system sizes are plotted (the legend is provided above the plots) and the black dotted line in the top right plot is an extrapolation to the thermodynamic limit. Bottom Row) Left and Middle - Derivatives of the Energy per site and average doublon occupancy with respect to  $\tau'$. Right - Scaling of $\langle S^{+}S^{-} \rangle$ with $L$ for several $\tau'$. Dotted Lines correspond to the polynomials $\langle S^{+}S^{-} \rangle = L^{2}/36 + 7L/18 + 10/9$, $\langle S^{+}S^{-} \rangle = L^{2}/36 - 5L/18 + 4/9$ and $\langle S^{+}S^{-} \rangle = 0$ for the respective $\tau'$. These polynomials have been used to determine the value of $\langle S^{+}S^{-} \rangle / L^{2}$ in the thermodynamic limit.}
\label{Fig:SF2}
\end{figure*} 

\begin{figure}
\centering
\includegraphics[width = 0.8\textwidth]{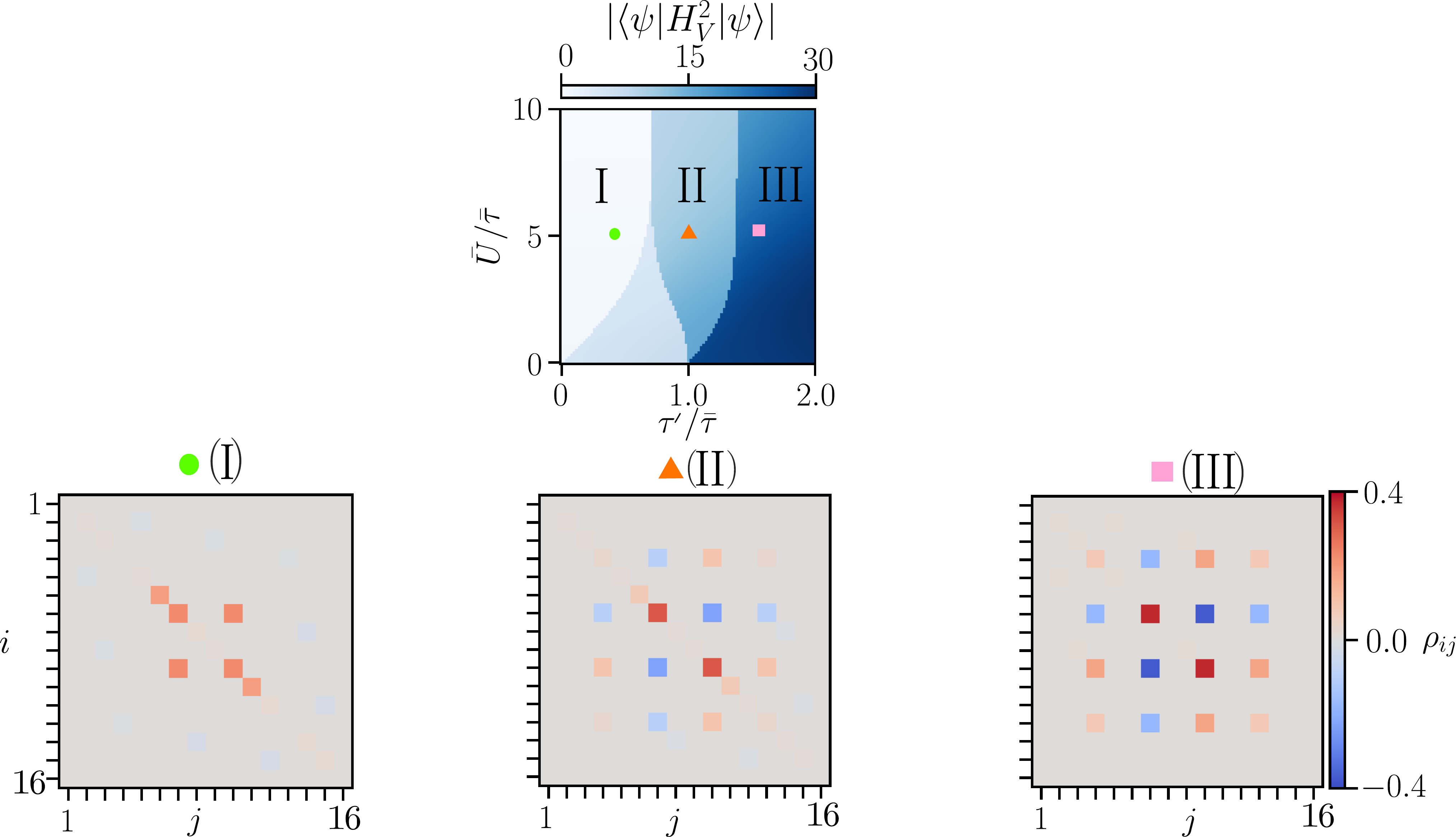}
\caption{Ground State Properties of the $L=8$-site two-rung triangular Hubbard model described in Eq. (\ref{Eq:SKappaHam}). Top row) Expectation value of the square of the vertical hopping integral, $H_{V}$, for a given $\bar{U}$ and $\tau'$. Second row) Average reduced density matrix for a pair of vertically bonded sites. We have used $U = 5.0\bar{\tau}$ and $\tau' = 0.5, 1.0$ and $1.5\bar{\tau}$ respectively. The indices $i$ and $j$ run through the different basis vectors with $\ket{\uparrow \downarrow, \uparrow \downarrow}$ corresponding to $i = 1$. The second quantum number changes each time the index increases by $1$ in the cyclic order $\ket{\uparrow \downarrow} \rightarrow \ket{\uparrow} \rightarrow \ket{\downarrow} \rightarrow \ket{0} \rightarrow \ket{\uparrow \downarrow} \hdots$ and the first quantum number changes, in the same order, every fourth increment.}
\label{Fig:SF3}
\end{figure} 

\begin{figure}
\centering
\includegraphics[width = 0.8\textwidth]{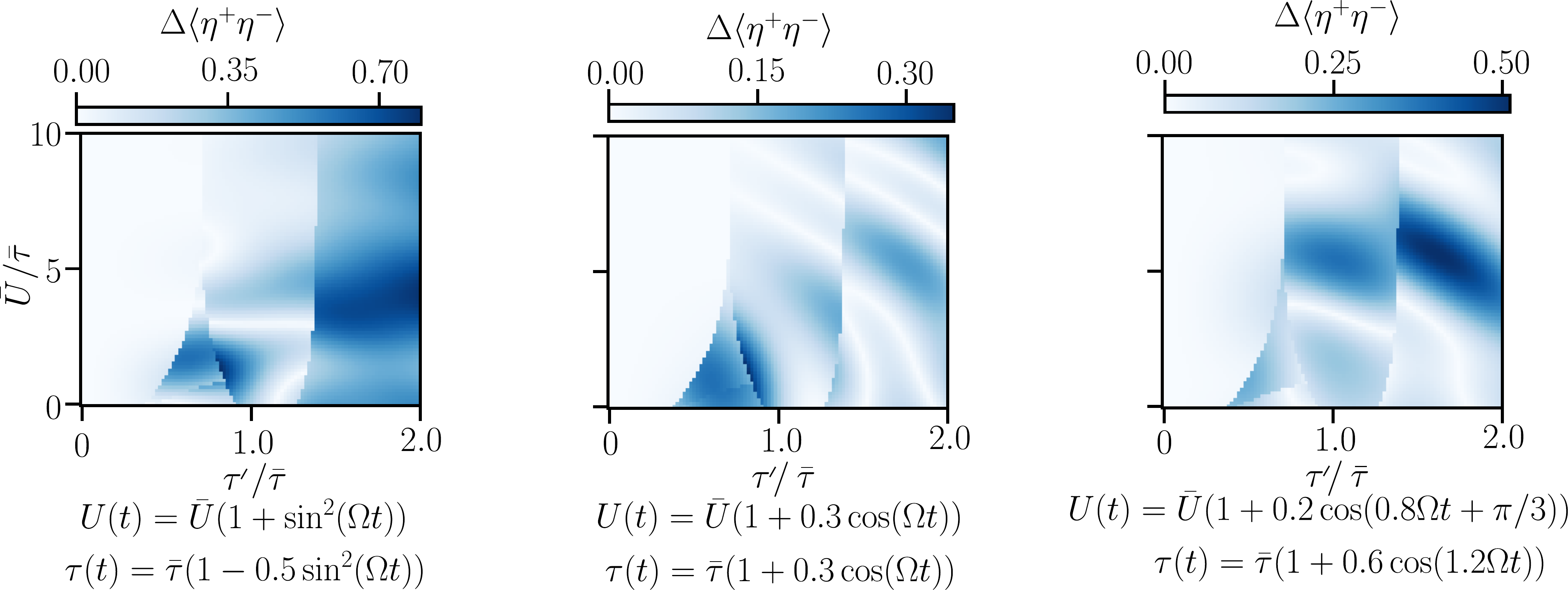}
\caption{Change in $\langle \eta^{+}\eta^{-} \rangle$ after time-evolving the ground state of the $L=8$ two-site triangular Hubbard model under $H(t)$ (see Eq. (\ref{Eq:SKappaHam})) with the specified $\bar{U}$ and $\tau'$. Each plot corresponds to a different time-dependence of the interaction strength and first hopping integral. We fix $\Omega = 2.25\bar{\tau}$ and calculate $\Delta \langle \eta^{+}\eta^{-} \rangle = \langle \eta^{+}\eta^{-} \rangle(t_{f}) - \langle \eta^{+}\eta^{-} \rangle(0)$, where $t_{f}\bar{\tau} = 2.0$}
\label{Fig:SF4}
\end{figure} 

\begin{figure}[h]
\centering
\includegraphics[width = 0.8\textwidth]{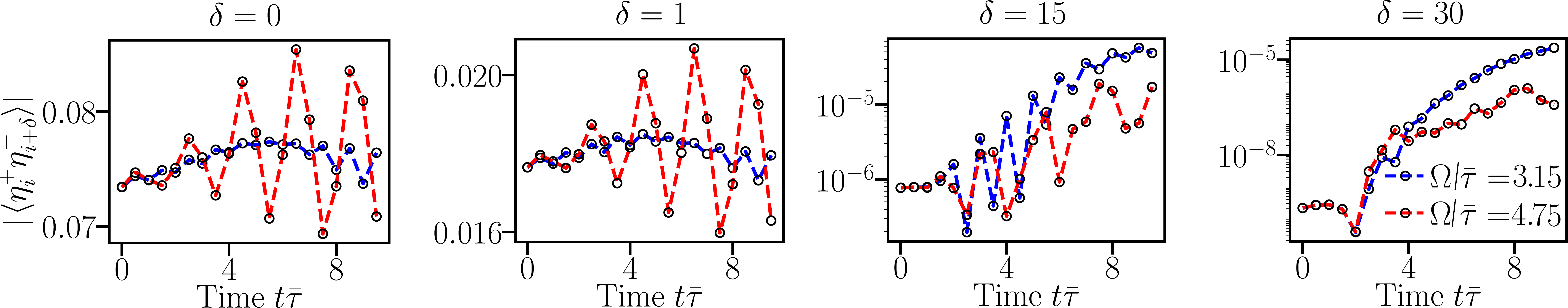}
\caption{Simulation of the half-filled driven $L = 32$ triangular Hubbard model in Eq. (\ref{Eq:SKappaHam}). The system is initialized in the ground state of $H(0)$, setting $\bar{U} = 4.82\bar{\tau}$ and $\tau' = 0.25\bar{\tau}$. The system is then time evolved under $H(t)$, using the same $\bar{U}$ and $\tau'$ and setting $A_{1} = -0.15 ,A_{2} = -0.075$, $T_{w} = 2.5\tau(0)$, $T_{p} = 5.0\tau(0)$ and $\Omega$ to the specified frequency (red, $\Omega / \bar{\tau} = 3.25$, blue $\Omega / \bar{\tau} = 4.75$). These frequencies corresponds to the peaks of Figs 4 in the main text. Here we plot the time evolution of the average magnitude of the particle-hole correlations at distances $\delta = 0, 1, 15, 30$ respectively. For $\delta = 0$ this corresponds to the on-site doublon density. For $\delta = 15$ and $\delta = 30$ the y=axis is plotted on a logarithmic scale.}
\label{Fig:SF5}
\end{figure} 

\end{document}